\documentclass[prl,aps,twocolumn,superscriptaddress,preprintnumbers,showpacs,floatfix,amsmath,amssymb]{revtex4}

\usepackage{bm}

\usepackage{amsmath}
\usepackage{graphicx}
\usepackage{amssymb}
\usepackage{mathrsfs}
\usepackage{bbm}

\newcommand{\be}{\begin{eqnarray}}
\newcommand{\ee}{\end{eqnarray}}

\begin{document}

%
%
\title{Topological Solitons and Folded Proteins}

\author{M.N. Chernodub}\thanks{On leave of absence from ITEP, Moscow, Russia}
\email{Chernodub@lmpt.univ-tours.fr }
\affiliation{Laboratoire de Math\'ematiques et Physique Th\'eorique,
Universit\'e Fran\c{c}ois-Rabelais Tours, F\'ed\'eration Denis Poisson - CNRS,
Parc de Grandmont, 37200 Tours, France}
\affiliation{Department of Mathematical Physics and Astronomy,
Krijgslaan 281, 59, Gent, B-9000, Belgium}
\author{Shuangwei Hu}
\email{Shuangwei.Hu@lmpt.univ-tours.fr}
\affiliation{Laboratoire de Math\'ematiques et Physique Th\'eorique,
Universit\'e Fran\c{c}ois-Rabelais Tours, F\'ed\'eration Denis Poisson - CNRS,
Parc de Grandmont, 37200 Tours, France}
\affiliation{Department of Physics and Astronomy, Uppsala University,
P.O. Box 803, S-75108, Uppsala, Sweden}
\author{Antti J. Niemi}
\email{Antti.Niemi@physics.uu.se}
\affiliation{Department of Physics and Astronomy, Uppsala University,
P.O. Box 803, S-75108, Uppsala, Sweden}
\affiliation{Laboratoire de Math\'ematiques et Physique Th\'eorique,
Universit\'e Fran\c{c}ois-Rabelais Tours, F\'ed\'eration Denis Poisson - CNRS,
Parc de Grandmont, 37200 Tours, France}

\begin{abstract}
We propose that protein loops can be interpreted as topological domain-wall solitons. They interpolate between
ground states that are the secondary structures like  $\alpha$-helices and $\beta$-strands.  Entire proteins can
then be folded simply by assembling  the solitons together, one after another. We present a simple
theoretical model  that realizes our proposal and apply it to a number of biologically active proteins
including 1VII, 2RB8, 3EBX (Protein Data Bank codes).  In  all the examples that we have considered
we are able to construct  solitons  that
reproduce secondary structural motifs such as $\alpha$-helix--loop--$\alpha$-helix and $\beta$-sheet--loop--$\beta$-sheet
with an overall root-mean-square-distance accuracy of around 0.7 \.Angstr\"om or less
for the central $\alpha$-carbons, {\it i.e.} within the limits of current experimental accuracy.
\end{abstract}


\date{\today}

\maketitle

Solitons are ubiquitous and widely studied objects that can be materialized in a variety of  practical
and theoretical scenarios \cite{sol1}, \cite{sol2}. For example solitons can be deployed for data transmission in transoceanic
cables, for conducting electricity in organic polymers \cite{sol1}, and they may also transport chemical energy in proteins \cite{dav}.
Solitons explain the Meissner effect in  superconductivity and dislocations in liquid crystals \cite{sol1}. They also model
hadronic particles, cosmic strings and magnetic monopoles in  high  energy physics \cite{sol1} and so on.  The first
soliton to be identified is the Wave of Translation that was observed by John Scott Russell in the Union Canal
of Scotland. This  wave can be accurately described by an exact soliton solution of the Korteweg-de Vries (KdV)
equation \cite{sol1}. At least in principle it can also be constructed in an atomary level simulation
where one accounts for each and every water molecule in  the Canal, together with all of their mutual interactions.
However,  in such a {\it Gedanken} simulation it would probably become a  real challenge to unravel the
collective excitations that combine into the Wave of Translation without any guidance from the known soliton
solution of the KdV equation since solitons can {\it not} be constructed simply by adding up small perturbations
around some ground state:  A (topological) soliton emerges when non-linear interactions  combine elementary
constituents into a localized collective excitation that is  stable against small perturbations and cannot decay, unwrap or
disentangle \cite{sol1}, \cite{sol2}.

In this Letter we propose  that  (topological) solitons can also explain and describe the folding of proteins into their native
state \cite{fold}, \cite{huang}.
We characterize a folded protein by the Cartesian
coordinates  $\mathbf r_i$ of its  $N$ central $\alpha$-carbons,
with $i=1,...,N$. For many biologically active proteins these coordinates can be downloaded from Protein Data Bank
(PDB) \cite{pdb}.  Alternatively, the protein
can  be described in terms of  its bond and torsion angles that  can be computed from the PDB data.
For this we introduce the tangent
vector $\mathbf t_i$ and the binormal vector~$\mathbf b_i$
\begin{equation}
\begin{matrix}
\mathbf t_i = \frac{ \mathbf r_{i+1} - \mathbf r_i}{ | \mathbf r_{i+1} - \mathbf r_i |}  \ \ \ \ \& \ \ \ \ \
\mathbf b_i =
\frac{ \mathbf t_{i-1} \times \mathbf t_i}{ | \mathbf t_{i-1} - \mathbf t_i |}
\end{matrix}
\label{tb}
\end{equation}
Together with the normal vector $\mathbf n_i = \mathbf b_i \times \mathbf t_i$ we then have  three vectors
that are subject to the discrete Frenet \cite{flory}
equation
\begin{equation}
\left( \begin{matrix} {\bf n}_{i+1} \\  {\bf b }_{i+1} \\ {\bf t}_{i+1} \end{matrix} \right)=
\exp\{ - \kappa_i \cdot T^2 \} \cdot \exp \{ - \tau_i \cdot T^3 \}
\left( \begin{matrix} {\bf n}_{i} \\  {\bf b }_{i} \\ {\bf t}_{i} \end{matrix} \right)
\label{frenet}
\end{equation}
Here $T^2$ and $T^3$ are two of the standard generators of three dimensional rotations, explicitely in terms of the
permutation tensor we have $(T^i)^{jk} = \epsilon^{ijk}$.
From (\ref{tb}), (\ref{frenet}) we can compute the bond angles $\kappa_i $ and the torsion angles $\tau_i $ using
PDB data for $\mathbf r_i$. Alternatively, if we know these angles we can compute the coordinates $\mathbf r_i$.
The common convention is to select  the range of these angles so that  $\kappa_i$ is positive. In the
continuum limit where (\ref{frenet}) becomes the standard  Frenet equation for a continuous curve,
$\kappa_i \to \kappa(x)$ then corresponds to local curvature.

As an example we consider the 35 residue
villin headpiece protein with PDB code 1VII that has been widely investigated, both theoretically and
experimentally  \cite{fold}. For example in the state of the art simulation \cite{fold2} succeeded in
producing its fold for a short time within an accuracy of  $\sim 2 - 3$\.A.

From the PDB data we compute the values of bond angles $\kappa_i$ and torsion
angles $\tau_i $ and the result is displayed in  Figure 1(a), where we use the
(standard) convention that the discrete Frenet curvature $\kappa$ is positive. In 1VII
there are three $\alpha$-helices that are separated by two loops.  When we use the PDB (NMR) convention
for indexing the residues the first, longer, loop is located at sites 49-54 and the second, shorter, between 59-62.

\begin{figure}[htb]
        \centering
                \includegraphics[width=0.45\textwidth]{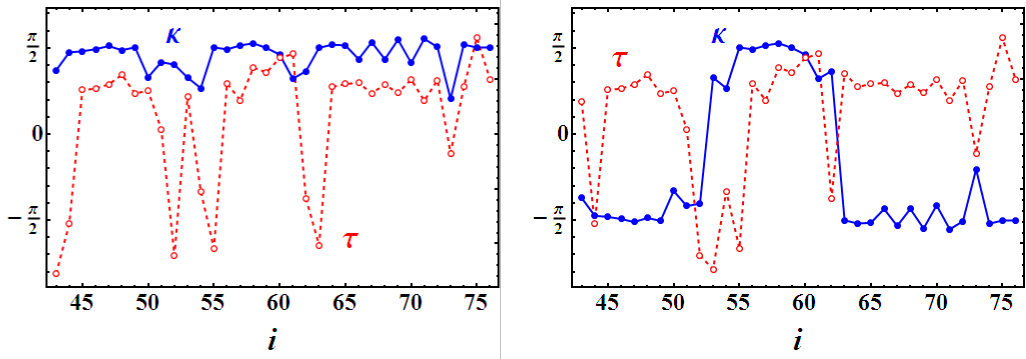} \newline
\hskip 31mm (a) \hskip 36mm (b)
        \caption{(a) The  bond and torsion angles of 1VII, computed
       with the (standard) convention that the discrete Frenet curvature $\kappa$ is positive.
       (b) The $\mathbb{Z}_2$ gauge transformed bond and torsion angles.
                }
       \label{Figure 1}
\end{figure}

We shall now show that Figure 1(a)  describes two soliton configurations, albeit in an encrypted form.
In order to decrypt the data in Figure 1(a) so that these solitons become unveiled
we observe that the equation (\ref{frenet}) has the following
local $\mathbb {Z}_2$ gauge symmetry: At every site we can send
\begin{equation}
\mathbb {Z}_2: \ \ \   \left\{ 
\begin{matrix} 
\kappa_i  & \to & \kappa_i \cdot \cos ( \Delta_{i+1} )\\
\tau_i    & \to & \tau_i+ \Delta_i - \Delta_{i+1} 
\end{matrix}  \right.
\end{equation}
and when  we choose at each site $\Delta_i = 0$ or $\Delta_i = \pi$
where $\Delta_i = \pi$ is the nontrivial element of the $\mathbb {Z}_2$ gauge group,
the Cartesian coordinates $\mathbf r_i$ computed from the discrete Frenet equation remain intact.
If we judiciously implement this $\mathbb {Z}_2$ gauge transformation in the data displayed in Figure 1(a) we arrive at the apparently
quite different Figure 1(b). Unlike  in Figure 1(a), the profile of $\kappa_i$ in Figure 1(b)  clearly displays the
hallmark profile of a topological soliton-(anti)soliton pair in a double-well potential: The two solitons are
located around the sites with indices 49-54 and 59-62 which are the locations of the  two loops in 1VII.  These solitons
interpolate between  the two "ground state" values  $\kappa_i \approx \pm \pi/2$
that pinpoint the locations of the  $\alpha$-helices in 1VII. Moreover, the two downswings in the value of $\tau_i$ from the value
$\tau_i \approx 1$ that mark  the locations of the $\alpha$-helices, coincide with  the locations of the two
solitons. The ensuing combined profile of $\kappa_i$ and $\tau_i$  is qualitatively consistent with a double-well
potential structure in the $(\kappa,\tau)$ plane that has the form displayed in Figure 2:
When we move from left to right in Figure 1(b), we follow a trajectory in the $(\kappa, \tau)$ plane that starts by fluctuating around the
potential energy minimum at $(\kappa,\tau) \approx (-\pi/2,1)$ in Figure 2,  corresponding to the first $\alpha$-helix. The
trajectory  then moves  through the first loop {\it a.k.a.} soliton (the red dashed line) to the second potential energy minimum {\it i.e.}
$\alpha$-helix   at $(\kappa,\tau) \approx (+\pi/2,1)$ in Figure 2, and  finally back through  the second loop {\it a.k.a.}
soliton (the blue solid line) to the first potential energy minimum at $(\kappa, \tau) = ( -\pi/2, 1)$.
\begin{figure}[htb]
        \centering
                \includegraphics[width=0.35\textwidth]{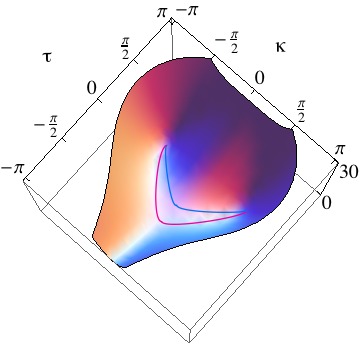}
        \caption{
     The potential energy on $(\kappa, \tau)$ plane that corresponds qualitatively to the data in Figure 1(b), the
     soliton between sites 49-54 corresponds to the red dashed trajectory and the soliton between sites 59-62 to the
     blue solid trajectory.}
       \label{Figure 2}
\end{figure}
We now present a simple theoretical model \cite{oma1}, \cite{oma2} that reproduces the $(\kappa,\tau)$ profile in Figure 1(b) as
a combination of two  soliton solutions,
with a very high atomary level accuracy for the central $\alpha$-carbons.
The model is defined by the  energy functional
\[
E = \sum\limits_{i=1}^{N-1} (\kappa_{i+1} - \kappa_i)^2 + \sum\limits_{i=1}^N
c\cdot (\kappa_i^2 - m^2)^2
\]
\vskip -2mm
\begin{equation}
+ \sum\limits_{i=1}^N \left\{ b \, \kappa_i^2 \tau_i^2 + d \, \tau_i + e \, \tau^2_i +
q \, \kappa^2_i \tau_i
\right\}
\label{E}
\end{equation}
Here $N$ is the num\-ber of central $\alpha$-carbons and $(c,m,b,d,e, q)$ are parameters.
The first sum describes nearest neighbor interactions along the protein.
The second sum describes a local self-interaction of the bond angles. The third sum describes local interactions
between bond and torsion angles, its  first term has an origin in a Higgs effect which is due to the potential term in the
second sum. The second term in the third sum is the Chern-Simons term, it is responsible for the chirality of the protein chain.
The third term is a Proca mass term and the last term can also be related to the Abelian Higgs Model, and it is also chiral.
As explained in \cite{oma2} this energy functional is essentially unique, and in particular it can be related
to a gauge invariant (supercurrent) version of the energy of 1+1 dimensional lattice Abelian Higgs Model. In three
space dimensions this model is also known as  the Ginzburg-Landau Model of conventional superconductivity \cite{sol2}.
Note that  in (\ref{E}) there is no reference to the  specifics of the interactions involving
amino acids such as  hydrophobic,  hydrophilic, long-range Coulomb, van der Waals, saturating hydrogen bonds {\it etc.}
interactions that are presumed to drive the folding process. The only explicit long-range force present in (\ref{E})
is the nearest neighbor interaction  described by the first term.
Moreover, as it stands (\ref{E}) depends only on {\it six}  site-independent, homogeneous parameters. There
is no direct reference whatsoever to the underlying in general highly inhomogeneous
amino acid structure of a protein.
We argue that this becomes possible  since (\ref{E}) supports {\it solitons} that describe
the common secondary structural motifs such as  $\alpha$-helix/$\beta$-strand - loop -  $\alpha$-helix/$\beta$-strand
as solutions to its  classical equations of motion.  Furthermore, even though the actual
numerical values of the parameters are certainly motif  dependent and for long loops that constitute bound states
of several solitons one might need to introduce
more than six parameters, we expect  there to be wide {\it universality} so that a given soliton with its relatively
few parameters describes a general class of homologous  motifs. Consequently  only a relatively small set of
parameters  are needed to provide soliton templates for structure prediction. In fact, we  propose that solitons are
the mathematical manifestation of the experimental observation, that the number of different protein folds is surprisingly limited.
The presence of solitons could then be the reason for the success of bioinformatics based homology modeling in
predicting native folds \cite{fold}.In order to quantitatively disclose the soliton solution of (\ref{E})  we start by observing that the first two sums
in (\ref{E}) can be interpreted as a discrete version of the energy of the 1+1 dimensional double well $\lambda \phi^4$  model that
is known to support the topological kink-soliton. In the continuum limit the kink has the analytic form \cite{sol1},
\cite{sol2},
\[
\kappa (x) = m  \cdot \tanh[ m\sqrt{c} \cdot ( x - x_0 ) ]\,.
\]
We  can try to estimate the parameters $m$ and $c$ for each of the two  solitons in the Figure 1(b) by a least square
fitting where we use this continuum soliton to approximate the exact soliton solution of the discrete equations of motion.
We consider here explicitly only the first soliton of 1VII, located between (PDB index) sites 49-54. Using the sites
46-56 we find the following least square fit
\begin{equation}
\kappa(x) \approx 1.4627 \cdot \tanh [ 2.0816 ( x - 52.597)] \,.
\label{k}
\end{equation}
In order to construct $\tau(x)$  we solve for its  equation of motion in (\ref{E}). The result is
\begin{equation}
\tau(x)  \approx  \ -2.4068 \cdot   \frac{1 - 0.4689 \cdot \kappa^2(x) }
{1 - 0.4619\cdot \kappa^2 (x) }
\label{t}
\end{equation}
In Figure 3 we show how the data in Figure 1(b) is described by the approximate soliton profile (\ref{k}), (\ref{t}).
\begin{figure}[htb]
        \centering
                \includegraphics[width=0.45\textwidth]{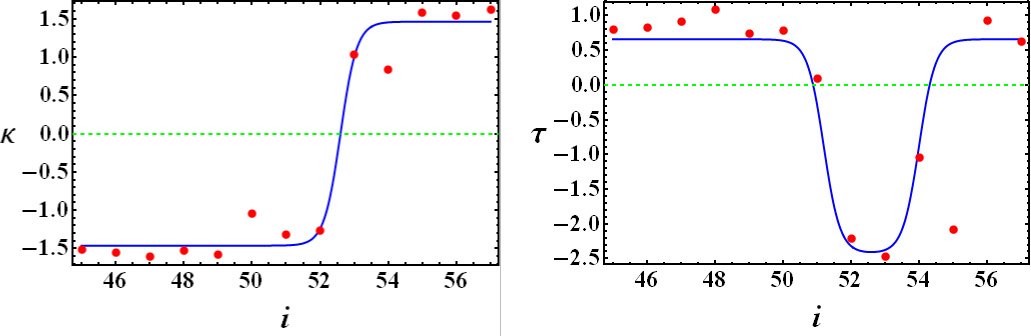}
        \caption{The PDB data for the first $\alpha$-helix - loop - $\alpha$-helix motif in 1VII, on the left $\kappa_i$
        and on the right $\tau_i$, together with the least square approximations (\ref{k}) and (\ref{t}) (the blue solid lines).}
       \label{Figure 3:}
\end{figure}
When we construct the ensuing discrete curve in the three dimensional space
by solving  (\ref{frenet}) with for $\kappa_i$ and $\tau_i$ given by  (\ref{k}) and (\ref{t}), we
reproduce the first loop of 1VII  with a surprisingly good RMSD
accuracy of $\sim$1.43 \.A  for the PDB indices 46-56 which is quite remarkable,
taking into account the simplicity of our approximation.

In order to construct a more accurate description of 1VII, we  resort to a numerical
construction of a soliton solution to the equations of motion if (\ref{E}). We use
simulated annealing that involves a Monte Carlo energy minimization
of the energy functional
\begin{eqnarray}
F & = & - \beta_1\cdot \sum\limits_{i=1}^N \Bigl\{  \left( \frac{ \partial E}{\partial \kappa_i} \right)^2  
+  \left( \frac{\partial E}{\partial \tau_i} \right)^2 \Bigr\}  
\label{F}\\
 & & - \beta_2 \cdot \sqrt{ \frac{1}{N} \sum\limits_{i=1}^N
| {\bf r}_{\mathrm{PDB}}(i) - {\bf r}_{\mathrm{soliton}}(i) |^2
}
\nonumber
\end{eqnarray}
with a simultaneous cooling of the two (inverse) temperatures $\beta_1$ and $\beta_2$.
Here the first sum vanishes when we have a solution to the classical difference equation of motion of (\ref{E}), and the
second sum  computes the RMSD distance between the $i$th $\alpha$-carbon of the solution and the protein we
wish to construct. The second term in (\ref{F}) acts like a chemical potential that selects the parameters in
(\ref{E}) so that we arrive at a soliton solution that corresponds to the given protein.

We have numerically constructed the classical solutions of (\ref{E}) that describe the secondary structural motifs in
proteins with PDB codes 1VII, 2RB8 and 3EBX. The first one has three $\alpha$-helices separated by loops, while the
second and third have $\beta$-strand-loop-$\beta$-strand motifs;  Both cases can be described equally by (\ref{E}),
the only difference  is that in the case of $\beta$-strands the two minima of the (classical) potential in (\ref{E})
are located at $(\kappa, \tau) \approx ( \pm 1, \pi)$. In each of the proteins that we have studied we
have routinely been able to  reproduce the secondary structural motifs as classical soliton solutions to the equations of motion
for (\ref{E})  in terms of only six parameters and with an overall RMSD accuracy of around  0.7 \.A per motif which is essentially the
experimental accuracy in X-ray crystallography and NMR; in our simulations
the first sum in (\ref{F}) decreases typically by around ten orders of magnitude indicating that the final configuration
is a solution, essentially within numerical accuracy.
Consequently  at least in these proteins the secondary structural motifs can be viewed as
solitons of the model (\ref{E}), within experimental accuracy. Since the
motifs that we have considered are quite generic in PDB data,  we have very little doubt that our results will
continue to persist whenever we have  loops  that
connect $\alpha$-helices and/or $\beta$-strands. And as long as the loops are not very long and do not describe bound states
of several solitons there does not appear to be any need
to introduce more than six parameters. Work is now in progress
to systematically construct and classify the solitons that describe the secondary structural motifs in a large class
of biologically active proteins.

We have also  made tentative  attempts to use our solitons to reconstruct entire proteins, by {\it naively} joining the solitons
that describe the secondary structural motifs at their ends. In the case of 1VII we have been able to reproduce in this
manner the entire protein as a classical soliton
with an overall RMSD accuracy of around 1.2 \.A and the result is
shown in Figure 4.
\begin{figure}[htb]
        \centering
                \includegraphics[width=0.45\textwidth]{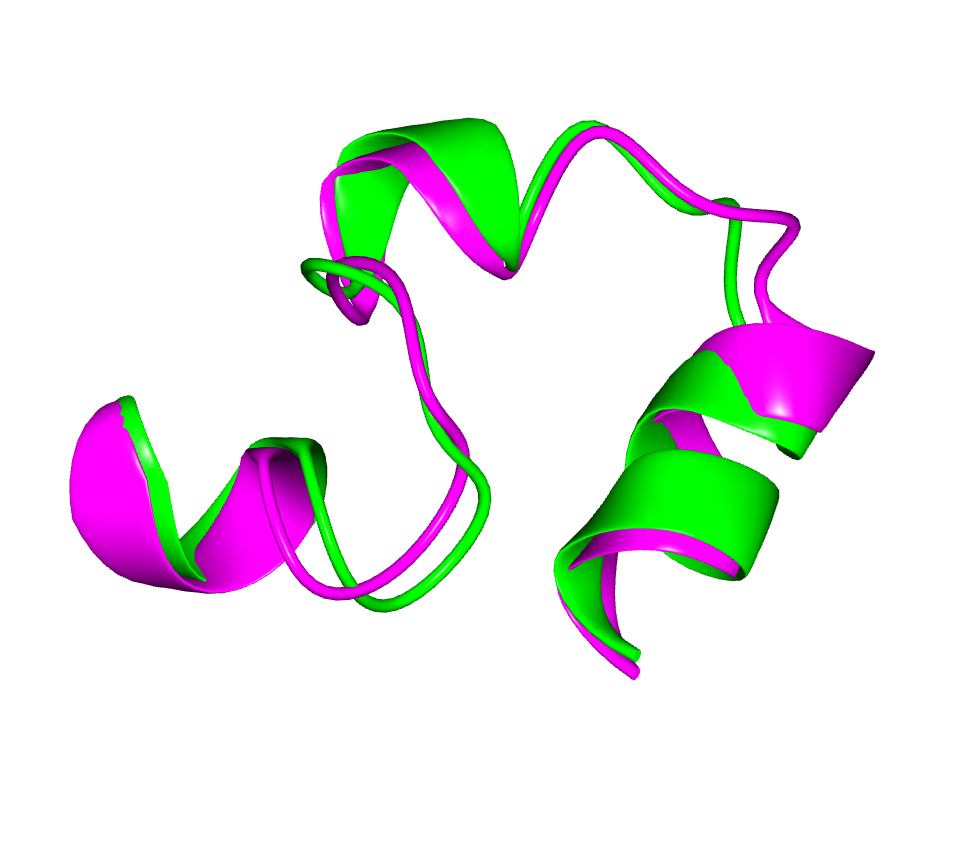}
        \caption{The helix-loop-helix-loop-helix structure of the 1VII protein (green) together with its reconstruction
        in terms of two solitons (purple). The RMSD distance between the two configurations is $\approx$ 1.2 \.A. }
       \label{Figure 4:}
\end{figure}
Even though the accuracy we obtain is very good,  the loss of accuracy from $\sim$ 0.7\.A to
$\sim$ 1.2 \.A when we combine the two solitons suggests that we can still substantially improve the method
of assembling an entire folded protein from its  solitons. Work is now in progress to develop more efficient methods
for assembling entire proteins from their solitons.

\vskip 0.3cm
In conclusion, we have proposed that the common secondary structural motifs that describe loops connecting
$\alpha$-helices and/or $\beta$-strands can be interpreted as topological solitons,  with the $\alpha$-helices
and $\beta$-sheets viewed as ground states that are interpolated by the loops as solitons.
Entire proteins can then be assembled simply by combining these solitons together one after another.
We have also presented a model that allows us to fold proteins in terms of its solitons within experimental accuracy.
In its simplest form that we have considered here,
the model has only six site independent but in
general motif dependent parameters. This appears to be sufficient to describe loops that are not too long.
This observation that all the details and complexities of amino acids and their interactions can be summarized
in so simple terms suggests the existence of wide universality in protein folding, and it can be viewed as a mathematically
precise formulation of the experimental observation that the number of protein conformations is far more
limited than the number of different amino acid combinations.  Finally, we leave it as a future challenge to expand the
model so that it incorporates an order parameter that describes the local orientation of the amino acids along the
$\alpha$-carbon backbone.

\vskip 0.2cm
Our research is supported by grants from the Swedish Research Council (VR). We thank
Martin Lundgren for discussions.

\end{document}